\documentclass[twocolumn,showpacs,preprintnumbers,amsmath,amssymb,eps]{revtex4}
\usepackage{graphicx}% Include figure files
\usepackage{dcolumn}% Align table columns on decimal point
\usepackage{bm}% bold math

\begin{document}

%\preprint{APS/123-QED}

\title{Electronic Phase Separation Transition as the Origin of the 
Superconductivity and the Pseudogap Phase of Cuprates\\}

\author{E. V. L de Mello, R. B. Kasal, Otton S. T. Filho and C. A. C. Passos}
%\altaffiliation[]{evandro@if.uff.br}
%\author{Raphael B. Kasal, C. A. Passos}
\affiliation{%
Instituto de F\'{\i}sica, Universidade Federal Fluminense, Niter\'oi, RJ 24210-340, Brazil\\}%
%\author{Otton Teixeira da Silveira Filho}
%\affiliation{Instituto de Computac\~ao, Universidade Federal Fluminense, Niter\'oi, RJ 24210-340, Brazil}

\date{\today}% It is always \today, today,

\begin{abstract}
We propose a new phase of matter, an electronic phase separation transition 
that starts near the upper
pseudogap and segregates the holes into high and low density domains. 
The resulting grain boundary potential favors the
development of intragrain superconducting amplitudes.
The zero resistivity transition arises only when the intergrain 
Josephson coupling $E_J$ is of the
order of the thermal energy and phase locking among the 
superconducting grains takes place. We show that this approach 
explains the pseudogap and superconducting phases in a natural way and reproduces
some recent scanning tunneling microscopy data.

\end{abstract}

%\keywords{Suggested keywords}%
\maketitle
%\section{Introduction}

The nature of the  pseudogap phase
has been widely recognized  to be a key for understanding the 
physics of cuprate superconductors  and its complex phase diagram\cite{TS,Tallon}.
At present there is no consensus on its origin
and also no agreement on the 
detailed generic doping dependence $p$
of the pseudogap temperature $T^*(p)$\cite{Tallon}. This difficulty to find an
explanation for the data collected by many different
experiments is certainly due to the intricate  charge 
dynamics of cuprate superconductors.

To deal with this complicate charge dynamics we have
proposed a static phase separation\cite{Mello03,Mello04,DDias07,DDias08} based on the
experimental evidences of ion diffusion in $La_2CuO_{4+\delta}$
and in Bi2212 above room temperature.  
The experimental signals that can be linked with a phase separation
are observed at the upper pseudogap $T^0(p)$ (in the notation of Ref.\cite{TS})
and consequently the ionic segregation transition 
must occur at a higher temperature $T_{PS}(p)$.
Since $T^0(p)$ ( and $T_{PS}(p)$) falls to zero in the overdoped
regime and ionic mobility requires high temperatures,
we assumed previously a charge disorder for underdoped compounds
and an uniform charge distribution for 
$p\ge0.20$\cite{DDias07,DDias08,JL07}. However, new scanning tunneling
microscopy (STM) data have shown an inhomogeneous local
gap structure that remains in the far overdoped 
regime\cite{McElroy,Gomes,Pasupathy} which cannot be
explained by an ionic phase separation, due to the low
values of $T^0(p)$ for large $p$.

These  STM results on different doping  regimes 
have clearly observed local gaps with different amplitudes
at temperatures below and above $T_c(p)$\cite{Gomes,Pasupathy}
that ruled out ionic phase separation as the sole origin of the
cuprate inhomogeneities.
In order to have an unified description of the STM data in the
overdoped and underdoped regions of the phase diagram, we
define a distinct  phase of matter, an electronic phase separation (EPS).
In this transition,
the electrons (or holes) generate bubbles 
as the temperature decreases below the onset temperature 
$T_{PS}(p)$ and freezes at lower temperatures.  

\begin{figure}[!ht]
%\begin{center}
\includegraphics[height=6cm]{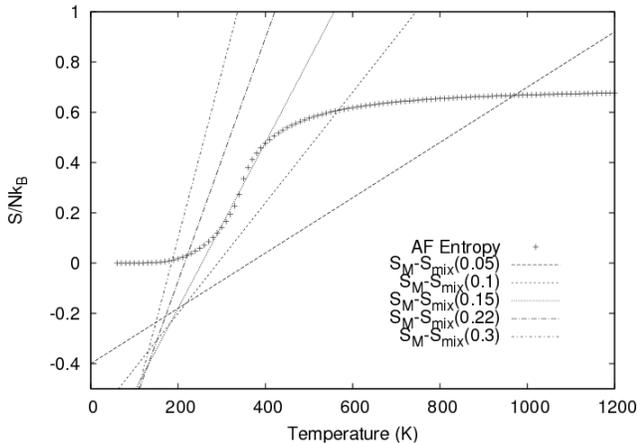}
\caption{ The lines are $S_{M}(p)-S_{Mix}(p)$ for some $p$
values. The intersections with AF entropy yields the onset of EPS, 
that is, $T_{PS}(p)$. } \label{EntBalFig}
%\end{center}
\end{figure}

The origin of this novel EPS transition is the
proximity to the insulator AF phase, common to all
cuprates, and can be described in terms of competing
minimum free energy or maximum entropy.
As the temperature decreases, the entropy of the
homogeneous density $p$ becomes lower than the anisotropic
system made of a granular bimodal distribution\cite{Mello03} of AF domains
with $p(i) \approx 0$ and high hole density domains with
$p(i)\approx 2p$.  This condition can be written as:
\begin{eqnarray}
S_{M}(p) \le +S^{2D}_{Is}+S_{Mix}(p).
\label{EntBal}
\end{eqnarray}
Where $S_{M}(p)=\gamma pT $ is the well known specific entropy of a 
homogeneous fermion gas with density $p$, $S^{2D}_{Is}$ is the Onsager
specific entropy for a 2D Ising model with spin coupling value that
yields a N\'eel temperature at $T=350$K, taken as a model to the
AF phase\cite{Stanley}. $S_{Mix}(p)$ is the entropy of mixing\cite{CH}. 
In  Fig.(\ref{EntBalFig}) we show the condition for
the EPS transition onset for some selected values of $p$, when the
straight lines cross the $S^{2D}_{Is}$. The used value of
$\gamma$ is consistent with the entropy 
measurements\cite{Loram}. The calculated
$T_{PS}(p)$ are in general agreement with the upper pseudogap
values\cite{TS,Tallon} $T^0(p)$  and, more importantly, {\it it provides
a physical interpretation for the origin of the electronic
inhomogeneities in the cuprates.}

\begin{figure}[!ht]
%\begin{center}
\includegraphics[height=6cm]{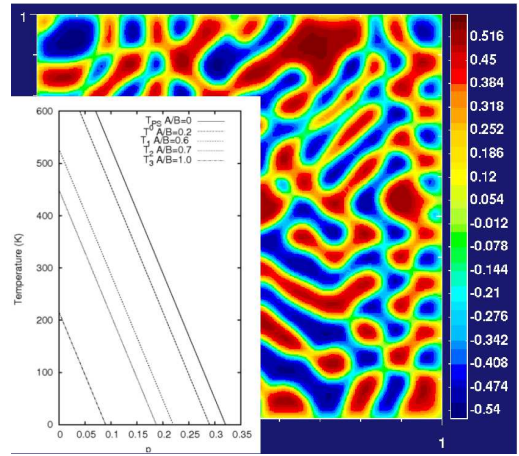}
\caption{(color on line) The charge density map on a $105 \times
105$ system after 6400 time steps and $A(T,p)/B=0.6$. The inset
shows estimates of $T_{PS}$, $T^0$ and some locus of
constant values of $A/B$.}
\label{Map}
%\end{center}
\end{figure}

Now that we have discussed why cuprates may
go through a transition to form granular 
charge domains,
we need to describe quantitatively such transition.
For this purpose we use
the theory of Cahn-Hilliard (CH)\cite{CH} that is appropriate
to describe a phase separation transition. The difference between the
local and the average  charge density  $u(i,T)\equiv (p(i,T)-p)$ 
is the order parameter.
Clearly $u(i,T)=0$ corresponds to
a homogeneous system above $T_{PS}(p)$. 
Then the typical Ginzburg-Landau free energy functional
in terms of such order parameter 
near the transition is given by

\begin{eqnarray}
f(i,T)= {{{1\over2}\varepsilon^2 |\nabla u(i,T)|^2 +V(u(i,T))}}.
\label{FE}
\end{eqnarray}
Where the potential ${\it V}(u,T)= A^2(T)u^2/2+B^2u^4/4+...$,
$A^2(T)=\alpha(T_{PS}(p)-T)$, $\alpha$ and $B$ are constants
that lead to lines of constant values of $A(T)/B$, parallel to
$T_{PS}(p)$, as shown 
in the inset of Fig.(\ref{Map}). $\varepsilon$ gives
the size of the grain boundaries among two low and high density
phases $p_{\pm}(i)$\cite{Otton,Mello04}. The energy barrier between
two grains of distinct phases is $E_g(T)=A^4(T)/B$ that is proportional to
$(T_{PS}-T)^2$ near the transition, and becomes nearly constant for
temperatures close to $T_{PS}(p)$. Thus, hereafter we will
use $E_g(p,T)\equiv V(p,T)$ as the grain boundary potential. 
$V(p,T)=V(p)\times V(T)$ and we assume, for simplicity, that
$V(p)$ have a linear behavior, whose equipotentials
are parallel to $T_{PS}(p)$. In the inset of Fig.(\ref{Map})
we plot $T_{PS}(p)$, $T^0(p)$ both assumed linear and the 
equipotentials are for $A(T_{PS}-T)/B$.

For completeness, the CH equation can be written\cite{Bray} in the
form of a continuity equation of the local free energy $f$,
$\partial_tu=-{\bf \nabla.J}$, with the current ${\bf J}=M{\bf
\nabla}(\delta f/ \delta u)$, where $M$ is the mobility or the
charge transport coefficient. Therefore,
\begin{eqnarray}
\frac{\partial u}{\partial t} = -M\nabla^2(\varepsilon^2\nabla^2u
+ A^2(T)u-B^2u^3).
\label{CH}
\end{eqnarray}
We have already made a detailed study of the density profile
evolution in a $105\times 105$ array as function of the time steps,
up to the stabilization of the local densities, for parameters that
yield stripe\cite{DDias07} and patchwork\cite{Mello04,Mello08}
patterns. 

The temperature evolution of the 
second order EPS  is studied by
the ratio $A(T)/B$\cite{Mello08}.   
$A(T)/B=0.2$ is close to the value of the measured upper pseudogap
temperature $T^0(p)$ shown in the inset of Fig.(\ref{Map}).
At $A(T)/B=0.6$, 
the EPS domains are clearly formed as displayed in Fig.(\ref{Map})
and the system is on the limit between a disordered metal with grains of
two densities, and a mixture of metallic and insulator (AF) grains.
This is possibly the origin of the instability that falls to
zero near p=0.18 as seen by may experiments\cite{Tallon,Loram,Hardy}
but not detected by the STM data for the Bi2212 series\cite{McElroy}.
At $T\approx 0$K the domains are 
frozen and in general the low density insulator regions
decrease in number and size
as $p$ increases, but even overdoped samples have
some remaining AF grains according to our simulations,
what is also experimentally verified by
the neutron diffraction data\cite{Tranquada}.

\begin{figure}[!ht]
%\begin{center}
\includegraphics[height=6cm]{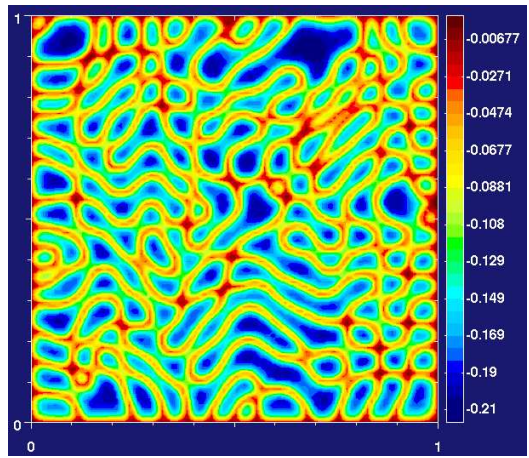}
\caption{(color on line) Local free energy (in arbitrary units)
density profile in the same location and temperature
as in Fig.(\ref{Map}). The dark (red) lines show the potential barrier
among the grain boundaries. } \label{MapFE}
%\end{center}
\end{figure}

We study the free energy evolution with
time and temperature together with the corresponding density profile.
In Fig.(\ref{MapFE}) we show the free energy map associated
with the density profile of Fig.(\ref{Map}), both made by the same 
computer simulation.
It shows that the
low and high density grains, at this temperature ($\%60$ of $T_{PS}$
for $p=0.16$),
are already bound regions of free energy minimum. 

%\section{The Local Gap}

As the temperature decrease below $T_{PS}$, the
potential barrier among the grains or the intragrain potential
$V(p,T)$ increases and becomes constant at low temperatures.
Consequently the holes become confined by
this effective  attraction toward 
the center of the grains and it may be taken as {\it the origin of the superconducting
interaction} that forms the (intragrain) hole pairs.

The intragrain superconductivity is naturally study 
with the Bogoliubov-deGennes (BdG) theory in a similar fashion
as we did before for a phenomenological 
potential and a static phase separation\cite{Mello04,DDias07,DDias08}. 
The calculations are performed  on a square
lattice of $32 \times 32$ sites, that is, on a small part of the 
charge density profile given in Fig.(\ref{Map}). 

Assuming the extended Hubbard Hamiltonian to describe
the  hole dynamics, the diagonalization
is made by the BdG 
equations\cite{Mello04,DDias07,DDias08,Mello08} with the hopping
value $t=0.15$eV, next neighbor hopping $t_2=0.70t$, on-site
repulsion $U=1.3t$ and, most importantly, the EPS next neighbor attraction
$V(p,T)$ derived from the values of $A(T)/B$.
Except from the temperature dependent $V(p,T)$, 
all the others parameters are similar to
values previously used\cite{Mello04,DDias07,DDias08}.

Following our free energy simulations, from low temperatures  
up to $T_{PS}(p)$ when the grains melt down, we can obtain
the qualitative behavior of $V(p,T)$. In order 
to yield average coherent gaps values comparable with
to the STM data on $0.11 \le p \le 0.19$
Bi2212 compounds\cite{McElroy}, we find  a set of parameters
that can be written as

\begin{eqnarray}
V(p,T)&=&V(p)\times V(T)=(-0.9+2.8 \times p)\times \nonumber \\
&& (1-T/T_{PS})^{(3-T/T_{PS})}, \label{VpT}
\end{eqnarray}
where the values are in $eV$, $V(p)$ is linear and vanishes
at $p\approx 0.32$ following
$T_{PS}(p)$. $V(T)$ falls to zero near $T_{PS}(p)$ and 
increases towards $T=0$K.

In general, the CH and BdG combined calculations yield very
low or almost zero local gaps for the regions with low densities,
that is, $p(i) \le 0.09$.
At the grains with larger local densities $p(i) \ge 0.1$, the local
Fermi level is large enough to have d-wave superconducting
amplitudes $\Delta_d(i,T)$. 
We define the {\it local superconducting temperature} $T_c(i)$
as the temperature which $\Delta_d(i,T)$ arises
in one given site "i". 
{\it The largest value of $T_c(i)$ in a given compound
determines the pseudogap temperature  
$T^*(p)$} which marks the onset of superconductivity. Since
$T^*(p)$ is close related with the potential $V(p,T)$ 
it also increases in the overdoped region, similar to 
the Nernst effect\cite{Ong} and many other experiments\cite{TS,Tallon}.

As the temperature decreases below $T^*(p)$ and some
high density grains become superconductors, 
the zero resistivity transition
takes place when the Josephson coupling $E_J$ among these
grains is sufficiently
large to overcome thermal fluctuations, that is, 
$E_J(p,T=T_c) \approx k_BT_c(p)$ what leads to phase locking
and long range phase coherence. 
Consequently the  superconducting
transition in cuprates occurs in two steps, similar to a superconducting material
embedded  in a non superconducting matrix\cite{Merchant}, first by
the appearing of intragrain superconductivity and by
Josephson coupling with phase locking at a lower temperature, {\it what provides
a clear interpretation to the pseudogap phase}.
Since $T_c(p)$ is not directly related with the local or intragrain
superconductivity, the gaps $\Delta_d(i,T)$ do not change appreciably
around $T_c(p)$, specially for underdoped compounds
that have large $T^*(p)$. This fact is verified experimentally by temperature
dependent tunneling\cite{Suzuki}
and angle resolved photon emission\cite{Campuzano}.

Using now the theory of
granular superconductors\cite{AB}, 
$E_J(p,T)\propto C_N(p) \times
\Delta(p,T)$ where $C_N(p)$ is the normal conductivity among
the grains. As shown in Fig.(\ref{Map}) the grain boundaries
are made of walls with the mean density $p$ surrounding 
the grains. On the other hand,
the conductivity increases a few
orders of magnitude with $p$ in the $Cu-O$ plane\cite{Takagi}
and $C_N(p)$ is small in the  underdoped region. That is just 
the opposite average behavior of $\Delta_d(i,T)$ that, following 
$V(p,T)$, decrease as $p$ increases.  This gives 
some insights on the superconducting
"dome shape" of the resistivity transition with
the maximum $T_c$ around $p=0.16$ in the middle of
the EPS region ($T_{PS}(p\approx 0.32)=0$). Also
$E_J \propto J_cr_i^2$, where $J_c$ is the critical current
density and $r_i$ is the average size of the grains. Taking typical
optimum doping values\cite{Tallon2}, that is, $J_c\approx
10^7A/m^2$ and $r_i\approx 50 \AA$ as one can see directly from
our Fig.(\ref{Map}), we get $E_J\approx 8$meV or
$T_c \approx 90$K, which is a good estimate for the Bi2212 optimum
$T_c$.

Now we turn to the new STM data that motivated the introduction
of the EPS concept. We firstly notice that 
the presence of the $p\approx 0$ AF insulator ($p(i) \le 0.03$)
and even low density domains ($p(i) \le 0.09$)
which are closer to the half filled band
and requires a high energy cost to accept extra electrons 
explains why injection of electrons produces less STM
current than extraction and also why this asymmetry increases drastically
as $p$ decreases\cite{Kohsaka}.
\begin{figure}[!ht]
\includegraphics[height=6cm]{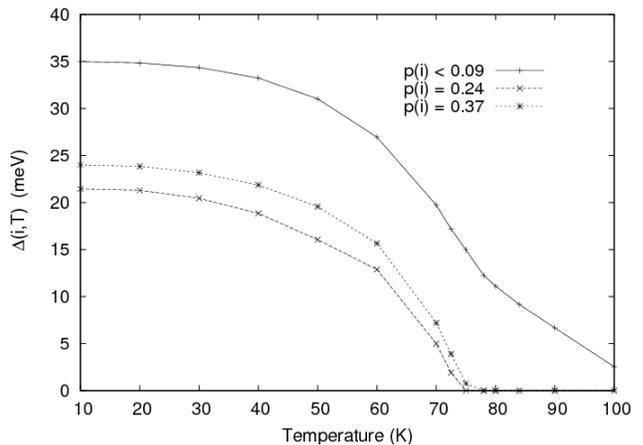}
\caption{The BdG calculation  for $\Delta_d(i,T)$ at 3 
locations on the $32 \times 32$ mesh with average hole doping $p\sim
0.24$. For $p(i) \le 0.9$ the STM signal is from activation over
the grain boundary potential and  
$\Delta_d(j,T)$. }
\label{MultGap}
\end{figure}

Fig.(\ref{MultGap}) shows some of the local BdG calculations
on selected points over a $Cu-O$ plane as in Fig.(\ref{Map}) to 
compare with the high  temperature STM data of
overdoped ($p \approx 0.22-0.24$) Bi2212 compounds\cite{Gomes,Pasupathy}. 
The smaller coherent gaps $\Delta_d(i,T)$ are from different
locations on metallic 
grains. The larger gaps originated in the insulator grains
and are due to activation over the grain boundary barrier $B_{ij}(T) \approx V(p,T)$
and the calculated $\Delta_d(j,T)$ from a metallic neighbor grain $j$.
As one can see in the experimental maps, at temperatures 
above $T_c(p)$, they are always 
surrounded by a small superconducting region\cite{Gomes,Pasupathy}.

An interesting consequence of this scenario is that {\it the 
lower density (insulator) grains have larger gaps but
lower local conductivity} that was verified experimentally\cite{Pasupathy}.
Another consequence is that, despite the uncertainty 
on $T_c(i)$ for very small gaps, the results 
follow close the measured relation
$2\Delta_d/K_BT_c(i) \approx 8$\cite{Gomes}.

\begin{figure}[!ht]
\includegraphics[height=6cm]{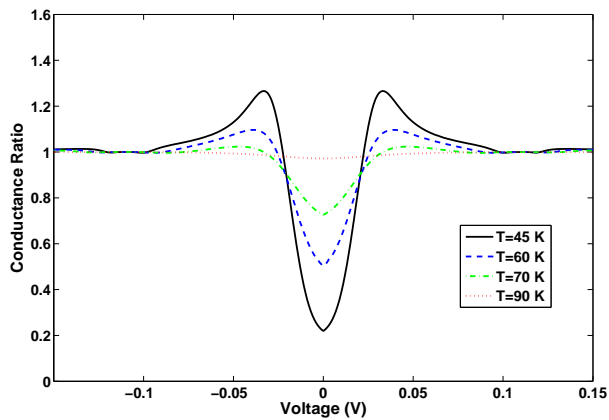}
\caption{(color online) The temperature dependent STM signal versus
applied voltage $V$ at a 
low density grain. The gap is due to 
activation energy over the potential barrier of
a superconducting metallic grain with a
local gap $\Delta_d(i,T)$.} 
\label{DE1}
\end{figure}

In Fig.(\ref{DE1}) we use our local gap calculation 
to plot the ratio between the tunneling conductance 
in the normal and in the superconducting state as measured
by Pasupathy et al\cite{Pasupathy}. Our calculation is
made with their Eq.(2). As mentioned above, 
for the case of insulator grains,
the total gap is the grain boundary barrier $B_{ij}(T)$ plus
the $\Delta_d(i,T)$ from a neighbor metallic grain. In Fig.(\ref{DE1})
we used $|\Delta_d|=25$meV and $B_{ij}(T)$=10mev similar to the larger
gap of Fig.(\ref{MultGap}).
Also, in  the
electronic granular scenario, due to the charge density oscillations
and charge tunneling through the grains,
the local inverse lifetime of the quasiparticle excitations\cite{Dynes}
$\Gamma$ is an oscillating  function of the applied bias $V$ and
the temperature. As
shown in Fig.(\ref{DE1}) this phenomenological form of $\Gamma$ yields the 
measured density of states\cite{Pasupathy} with a structure
near the applied voltage $V=0.1$eV that is captured by our calculations.

%The  local density of states
%was probed by studying the ratio between the tunneling conductance
%in the normal and in the superconducting state,
%namely\cite{Pasupathy}.
%\begin{eqnarray}
%&& \frac{N_d(i,U,T)}{N_N(i,T)}=\frac{1}{\pi}\int dE\frac{df(E+V,T)}{dE}
% \int_0^\pi d\theta Real \times \nonumber \\ &&
%\frac{E-i\Gamma(i,T)}{\sqrt((E-i\Gamma(i,T))^2-B_{ij}(T)^2-\Delta(i,T)^2cos^2(2\theta))}
%\label{Dpg}
%\end{eqnarray}
%Where $N_N(i,T)$ is the normal local density of states. $V$ is the
%applied bias. $N_d(i,U,T)$ is the local density of states of at "i"
%where $\Delta_d(i,T)$ is the coherent d wave local superconducting
%BdG gap. For the grains with $p(i) \le 0.09$, the total gap is the
%average of neighbors $\Delta_d(i,T)$  with the grain boundary
%barrier $B_{ij}$. $\Gamma(i,T)$ is the local inverse lifetime of the
%quasiparticle excitation, and in conventional superconductors
%$\Gamma$ depends only on the temperature\cite{Dynes}. In the
%electronic granular scenario, due to the density oscillations,
%$\Gamma$ is an oscillating function of the applied bias $V$. As
%shown in Fig.(\ref{DE1}) this form of $\Gamma$ yields the density of
%states measured by Pasupathy et al \cite{Pasupathy} with a structure
%near $V=0.1$eV that is captured by our calculations.

%\section{Discussions and Conclusion}
In conclusion we have proposed a new electronic phase to cuprate
superconductors
essentially made of disordered low and high charge density grains. 
The grains are static at low temperatures 
but melts slowly and disappear near $T_{PS}(p)$. 
Such anomalous phase arises
due to the proximity of the undoped insulator with AF order, and may be common
to other materials with some doping dependent phases, like manganites
which possess also a pseudogap phase\cite{Dagotto}.

%\section{Acknowledgment}
We gratefully acknowledge partial financial aid from Brazilian
agency CNPq.

\end{document}